\documentstyle[prbbib,aps,twocolumn]{revtex}
\input{psfig}
\title{Kinetics of Phase ordering in Microemulsions and Micellar Solutions}
\author{Debashish Chowdhury and Prabal K. Maiti}
\address{Physics Department, Indian Institute of Technology,\\
Kanpur 208016, India}
\begin{document}
\maketitle
We review the models developed and techniques used in recent years 
to study the kinetics of phase ordering in a class of complex fluids, 
namely, ternary microemulsions and micellar solutions.  

\section{Introduction}

The ``head'' part of surfactant molecules consist of a polar or ionic group. 
The ``tail'' of many surfactants consist of a single hydrocarbon chain 
whereas that of some other surfactants, e.g., phospholipids, are made 
of two hydrocarbon chains both of which are connected to the same head 
\cite{tanford}. In contrast, {\it gemini} surfactants 
\cite{Deinega,Menger1,Menger2,Rosen1}, 
consist of two single-chain surfactants whose 
heads are connected by a ``spacer'' chain and, hence, these ``double-headed'' 
surfactants are sometimes also referred to as ``dimeric surfactants'' 
\cite{Zana1,Zana2} (see fig.1). 
\begin{figure}
\centerline{\psfig{file=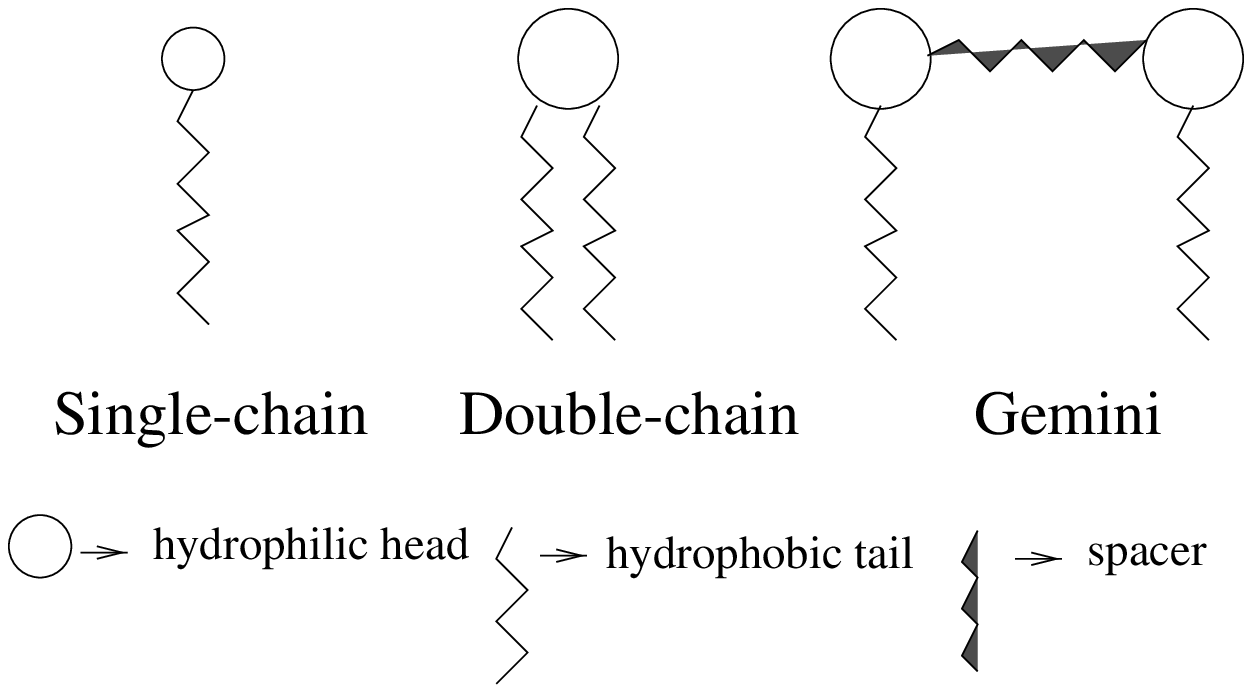,width=12cm}}
\caption{\sf Different types of amphiphiles.}
\label{fig1.1}
\end{figure}

When put into an aqueous medium, the ``heads'' of the surfactants like to get 
immersed in water and, hence, are called ``hydrophilic'' while the tails tend to 
minimize contact with water and, hence, are called ``hydrophobic'' \cite
{tanford}. The spacer in gemini surfactants is usually hydrophobic but gemini 
surfactants with hydrophilic spacers have also been synthesized \cite{Rosen2}. 
Surfactant molecules are called ``amphiphilic'' because their heads are 
``water-loving'' and hydrocarbon chains are ``water-hating''.
Because of their amphiphilicity the surfactant molecules form ``{\it 
self-assemblies}''  (i.e., supra-molecular aggregates), such as monolayer 
and bilayer membranes, micelles, inverted-micelles, etc.\cite{Gelbert},  
in a multi-component fluid mixture containing water. 
These not only find wide ranging applications in detergent and 
pharmaceutical industries, food technology, petroleum recovery, etc. 
but are also one of the most important constituents of cells in living 
systems. Therefore, physics, chemistry, biology and technology meet at 
at the frontier area of interdisciplinary research on association 
colloids formed by surfactants \cite{evans}.

A ternary microemulsion is a ``colloidal'' complex fluid consisting of 
three components, namely, oil, water and surfactants;  
the phase diagrams of such ternary systems have been studied 
extensively \cite{chen,Gelbert}.
When the concentration of the amphiphilic molecules 
is sufficiently high, the low-temperature equilibrium phase of the system 
is lamellar where the amphiphiles arrange themselves in (approximately) 
parallel stacks. On the other hand, when the concentration of the 
amphiphiles is not high, the system exhibits either a droplet phase or a 
bicontinuous microemulsion phase in equilibrium at sufficiently low 
temperatures depending on the relative concentrations of oil and water. 
If the concentrations of oil and water are comparable then the bicontinuous 
microemulsion phase is observed. But, if the concentration of oil (water) 
is much less than that of water (oil) then droplets of oil (water) are 
found to be dispersed in water (oil); these droplets are often referred 
to as micelles and, therefore, the system under such conditions are 
called micellar solution. These systems find wide ranging industrial 
applications.

The aim of this chapter is to present a systematic and upto date review 
of the models developed and techniques used so far to study the kinetics 
of phase ordering in microemulsions and micellar solutions (see references 
\cite{chow1,chow3,kawasaki4,larad} for earlier reviews).

\section{Kinetics of Ordering in Binary Mixtures in the Absence 
of Surfactant Impurities} 

We introduce the key concepts involved in the studies of phase 
ordering by illustrating them with the help of binary mixtures.
The techniques used so far for the studies of phase ordering in 
ternary microemulsions and micellar solutions are extensions of 
those used widely for the corresponding studies in the simpler 
case of binary mixture of immiscible components, e.g., binary 
alloys, binary fluids, etc. Therefore in this section we 
classify and summarize the main techniques used for such studies
\cite{gunton,Gusan,komura,binder,lang,bray}. 

Suppose a binary alloy has been quenched from a high temperature $T_h$ 
to a temperature $T_{\ell}$ well below the coexistence curve. The 
morphology of the coarsening domain ``pattern'' depends on the relative 
concentrations of the two components. If the concentrations of the
two components are 
comparable then the pattern has a random ``interconnected'' structure. 
On the other hand, if the concentration of one of the components is 
much smaller than that of the other the pattern consists of ``droplets'' 
of various sizes.

There are essentially two different (albeit complementary) theoretical 
approaches to the study of the kinetics of the phase ordering processes, 
viz., microscopic models and phenomenological models. 

\subsection{Microscopic Models} 

Two types of microscopic models have been used; the molecular models 
defined on a continuum and lattice models. 

\subsubsection{Molecular Models on a Continuum:} 
For the Molecular Dynamics simulation of an immiscible binary
mixture of two species, labelled $A$ and $B$, one first
postulates an approximate form of inter-molecular interactions,
e.g., truncated Lennard-Jones potentials. The potentials are
chosen in such a manner that the interactions $A-A$ and $B-B$
are attractive while the interaction $A-B$ is repulsive \cite{lepto}.
Normally one uses the Verlet algorithm for integrating
Newton's equation of motion \cite{rapport}. In principle,
one can use either a constant-energy ensemble or a constant
temperature ensemble.

\subsubsection{Microscopic Lattice Models}
The binary alloys can be modelled as an Ising spin system where
$S_i=1$ correspond to an $A$ atom and $S_i=-1$ correspond to an $B$
atom. In the symmetric case, i.e., when $E_{AA} = E_{BB}$, the Hamiltonian
for the system is given by 
\begin{equation}
H = - J\sum S_iS_j
\end{equation}
where $J > 0$ and the summation is to be carried out over all the distinct 
nearest-neighbour spin pairs. However, no such model is complete
without specification of the prescription for the dynamical evolution
of the system \cite{kawasaki72}. In the case of binary alloys, unlike
the magnetic counterpart, the order parameter is conserved, i.e., the
concentrations of $A$ atoms and $B$ atoms remains unaltered during the
time evolution in a closed system. The kinetics of ordering in such a 
system can be studied
at a microscopic level by using the so-called Kawasaki spin-exchange
dynamics: two anti-parallel nearest-neighbour spins can exchange
their position with a probability $1/[1+\exp(\beta\Delta E)]$, with
$\beta = 1/(k_BT)$ where $k_B$ is the Boltzmann constant and $\Delta E$
is the energy change that would be caused by the interchange of the two
spins. 
 
\subsection{Phenomenological Models with Langevin Dynamics}

In the phenomenological approach the system is described by an order parameter 
field $\psi(\vec r)$, which is the local difference in the concentrations of 
the $A$ and $B$ atoms. In contrast to the discrete allowed values of the spins 
in the lattice model, the order parameter can take all real values in the 
interval $-1 \leq \psi \leq 1$. The coarse-grained free-energy functional 
for the $d$-dimensional system is given by 
\begin{equation}
F[\psi(\vec r)] = \int d^dr [r_0 \psi^2(\vec r) + u \psi^4(\vec r) + 
c |\nabla \psi(\vec r)|^2] 
\end{equation}
where $r_0, u$ and $c$ are phenomenological coefficients. The symmetry 
requirements rule out the possibility of $\psi$ and $\psi^3$ terms in this 
functional. The dynamics of the system is assumed to be governed by the 
so-called Langevin equation 
\begin{equation}
\partial \psi(\vec r,t)/\partial t = \Gamma \nabla^2[\delta F/\delta 
\psi(\vec r, t)] + \eta(\vec r, t) 
\end{equation}
where $\Gamma$ is the phenomenological kinetic coefficient, 
$\delta/\delta \psi$ denotes the functional derivative with respect to 
$\psi$ and the Laplacian takes care of the fact that the order parameter is 
conserved. $\eta(\vec r,t)$ is the noise, which is usually assumed to be of 
`Gaussian white' nature, i.e.,
\begin{equation}
<\eta(\vec r,t) \eta(\vec r^{\prime},t')> = 2 k_B T \nabla^2 \delta(\vec r 
- \vec r^{\prime}) \delta(t - t') 
\end{equation}
where $k_BT$ guarantees the approach to the true Gibbsian equilibrium. 
In the context of the binary alloys the equation (3) is called the 
Cahn-Hilliard-Cook equation for historical reasons. The model described 
by the equations (2) - (4) is usually referred to as the model $B$ 
\cite{hohen}. The computation becomes much more efficient 
(i.e., the morphological characteristic of the asymptotic regime can be 
obtained with a very short computer time) by solving equation (3) above by 
the cell dynamics method \cite{ono}. 

\subsection{Characteristic Quantities of Interest} 

Two quantities are most important in describing the kinetics of this 
growth process, viz., the time dependence of the typical linear 
size $R(t)$ of the ordered regions and the dynamic scaling of the 
structure factor ${\cal S}(\vec q,t)$ characterizing the statistical 
self-similarity of the coarsening pattern at different times. The 
structure factor is the 
Fourier transform of the two-point correlation function in real space, i.e., 
\begin{equation}
{\cal S}(\vec q,t) = \sum_r G(\vec r,t) exp(i \vec q {\bf .} \vec r)
\end{equation}
where $G(\vec r,t) = <S(\vec 0,t) S(\vec r,t)>-<S>^2$ is the correlation 
function in real space. However, since numerical computations 
are carried out either on discrete lattices (in the case of microscopic 
models) or on discrete grids (in the case of phenomenological models) the 
corresponding structure factor is given by 
\begin{equation}
{\cal S}(\vec q,t) = <|(1/N) \sum_{\vec r_i} G(\vec r_i,t) 
exp(i \vec q{\bf .} \vec r_i)|^2> 
\end{equation}
with $\vec q = (2\pi/L)(mx + ny+pz)$ and $m,n,p = 1,2,3,...,L$. 

Following a rapid quench from a very high temperature to a temperature 
below the coexistence curve, coarsening of ordered domains takes place 
and, consequently, the first zero crossing of $G(R,t)$ (i.e., the 
smallest $R$ for which $G(R,t)=0$) occurs at larger and larger values of $R$ 
at successively longer values of time $t$, where $G(R,t)$ is the circularly 
averaged two-point correlation function. As a result, the location of 
the first zero crossing of $G(R,t)$ may be taken as a measure of $R(t)$. 

During the coarsening process the position of the peak in ${\cal S}(q,t)$ 
keeps moving towards smaller values of $q$ where ${\cal S}(q,t)$ is 
circularly averaged. The dynamical scaling form of the structure factor 
is given by 
\begin{equation}
{\cal S}(q,t) = R^d F(qR(t)) 
\end{equation}
where $F(x)$ is a function of $x = qR(t)$. The length $R(t)$ can be extracted 
from ${\cal S}(q,t)$ in several different ways:\\
(i) $R^{-1}(t) = k_m/2\pi$, where $k_m$ is the location of the maximum of the 
structure factor ${\cal S}(q,t)$, (ii) $2\pi/R$ may be identified with the 
first moment of the structure factor or the square root of the second 
moment of the structure factor, etc. 
These measures of $R(t)$ can also be used in the case of ternary 
microemulsions and micellar solutions. Moreover, one can compute 
the mean-cluster size of a particular component, say $A$, using 
the definition 
\begin{equation}
\chi_A = [\sum_{s=1}^{s_{max}-1} s^2 n(s)]/[\sum_{s=1}^{s_{max}} s n(s] 
\end{equation}
where $n(s)$ is the number of clusters of type $A$ with $s$ sites and 
$s_{max}$ is the corresponding size of the largest cluster. 

For binary alloys, for which the order parameter is conserved, i.e., for 
model $B$, the length $R(t)$ has been found to follow the growth law 
\begin{equation}
R(t) \sim t^n 
\end{equation}
where $n = 1/3$. In contrast, for systems with non-conserved scalar order 
parameter (model $A$) $n = 1/2$. Moreover, in case of binary fluids 
hydrodynamic effects lead to the growth law (9) with $n = 1$. 
Furthermore, 
\begin{equation}
R(t) \sim (log t)^x, 
\end{equation}
with dimensionality-dependent exponent $x$,
in phase separating binary systems in the presence of quenched 
disorder (e.g., impurities) \cite{chow2}.

\section{Kinetics of Ordering in Ternary Microemulsions and Micellar Solutions}

In our discussion of the kinetics of phase ordering in ternary microemulsions 
and micellar solutions we shall begin with the molecular models, which are 
truly microscopic description of the system, and then consider models at a  
coarser level and, finally, consider phenomenological models. 

\subsection{Molecular Dynamics Simulations of Molecular Models on a Continuum} 

Laradji et {\it al.} \cite{larad1}
have carried out a MD simulation of phase separation in a 
binary mixture in the presence of surfactants. The interaction potential 
between two molecules $i$ and $j$ (oil-oil, water-water or oil-water) is 
assumed to be 
\begin{equation}
U(\vec{r}_{ij}) = 4 \epsilon [(\sigma/r_{ij})^{12} - 
(2 \delta_{\alpha_i \alpha_j} - 1) (\sigma/r_{ij})^6] 
\end{equation}
where $\delta_{\alpha_i \alpha_j}$ is the Kronecker delta function and 
$\alpha_i$ denotes the type of the molecule $i$, i.e., $\alpha_i = 1$ 
for a water molecule and $2$ for an oil molecule. Thus, the potential 
above ensures that molecules of different species always interact 
repulsively. For example, if $\alpha_i \neq \alpha_j$, 
\begin{equation}
U(\vec{r}_{ij}) = 4 \epsilon [(\sigma/r_{ij})^{12} + (\sigma/r_{ij})^6] 
\end{equation}
On the other hand, if $\alpha_i = \alpha_j$, 
\begin{equation}
U(\vec{r}_{ij}) = 4 \epsilon [(\sigma/r_{ij})^{12} - (\sigma/r_{ij})^6] 
\end{equation}
In this model the ``water-loving'' head of every surfactant molecule is 
treated as identical to a water molecule and the ``oil-loving'' tail is 
treated as identical to an oil molecule. In other words, every surfactant
molecule is assumed to be, effectively, a diatomic molecule 
one part of which is water-like and the other part is oil-like; the two 
parts of the molecule are assumed to be connected to each other by a 
harmonic spring so that the potential is given by  
\begin{equation}
U_{ss}(\vec{r}) = (K_s/2)(r_{ij} - \ell_s)^2. 
\end{equation}
To our knowledge, this type of models were first proposed by  
Smit et {\it al.} \cite{smit}.
However, Laradji et {\it al.} \cite{larad1}
 was the first to study the kinetics of 
phase ordering by MD simulation of such a model. 
A non-zero concentration of amphiphiles was found to slow down the 
coarsening process leading to a significant deviation from the 
power-law (9). 

\subsection{Microscopic Lattice Models} 

\subsubsection{Widom Model: a Spin-1/2 Ising Model on Simple-Cubic Lattice 
with Farther-Neighbour Interactions} 

Widom model \cite{widom}
 is the simplest lattice model of ternary mixtures of oil, 
water and surfactants. This is a lattice model in the same spirit as, 
for example, the lattice model of binary alloys. However, in contrast to 
the lattice model of binary alloys described in the preceding section, 
the molecules of oil, water and surfactants in this model are located on 
the bonds rather than on the lattice sites. In this model the molecules 
of oil, water and surfactants are represented by the nearest-neighbour 
bonds of a spin-1/2 Ising model where every spin interacts with the  
nearest-neighbour as well as a specific subset of farther-neighbour spins. 
The bonds between the up-up nearest-neighbour spin pairs represent water 
molecules, those between down-down nearest-neighbour spin pairs represent 
oil molecules and those between antiparallel nearest-neighbour spin pairs 
denote the surfactant molecules. The Hamiltonian for this model is given by 
\begin{equation}
{\cal H} = - J \sum_{<ij>} S_i S_j - 2 M \sum_{<ik>} S_i S_k - 
M \sum_{<il>} S_i S_l 
\end{equation}
where, for three dimensional systems, the summations in the first, second 
and third terms on the right hand side are to be carried out over the 
nearest-neighbour, second-neighbour and fourth-neighbour spin pairs, 
respectively, on a simple cubic lattice. The interaction $J$ is positive 
(ferromagnetic in the terminology of spin models of magnetic materials) 
whereas $M$ is negative (antiferromagnetic). The farther-neighbour 
interactions arise from a prescription, suggested by Widom \cite{widom}
, to take 
into account the bending rigidity of the amphiphilic monolayer membrane 
at the oil-water interface. This model reduces to the standard spin-1/2 
Ising model with only nearest-neighbour ferromagnetic interactions on a 
simple cubic lattice if $M = 0$. 

For studying the dynamics of the Widom model Morawietz et {\it al.}
\cite{mora}
introduced a Kawasaki-type spin exchange dynamics which conserves the 
molecules of oil, water and surfactants. Note that in this algorithm 
the spins on the lattice sites, rather than the molecules on bonds, are 
exchanged; however, the algorithm is such that it leads to simultaneous 
exchange of three molecules while satisfying the condition that  
the number of molecules of 
each species remains conserved. The algorithm is as follows: 
a up-spin can exchange its position with one of its down-spin neighbours, 
with probability $1/(1+exp(\beta E))$, provided the numbers of up and 
down spins among the neighbours of the first spin (excluding the second 
spin, which is also a neighbour of the first) are the same as the 
neighbours of the second spin (excluding the first spin, which is also 
a neighbour of the second). 

Using this new algorithm, Morawietz et {\it al.}
\cite{mora} investigated the kinetics 
of dis-ordering, rather than the traditional study of ordering, by 
reverse quenching the system from the ordered phase to the disordered 
phase. More specifically, an initial configuration was created where 
a layer of surfactants in the central part of the system is sandwiched 
between a column of oil above and a column of water below. The parameters 
$J/(k_BT)$ and $M/(k_BT)$ were so chosen that the corresponding equilibrium 
phase is known to be a disordered fluid. To our knowledge, the kinetics of 
ordering in the Widom model has never been studied.

\subsubsection{Kawakatsu-Kawasaki Model: a Decorated Spin-1/2 Ising Model} 

The Alexander model \cite{alexander}
 is a decorated spin-1/2 model, which was extended 
by Chen et {\it al.} \cite{chen1} for a more realistic description 
of ternary microemulsions. 
Kawakatsu and Kawasaki \cite{kawaktsu}
 have studied the kinetics of phase ordering 
using a decorated spin-1/2 model which may be regarded as a simplified 
special case of the model considered by Chen et {\it al.}\cite{chen1}. 
In this model every 
lattice site is occupied by a Ising spin; $S_i = 1$ corresponds to a 
water molecule and $S_i = -1$ corresponds to a oil molecule at the $i$-th 
site. Just as in the molecular models mentioned in the preceding section, 
the ``water-loving'' head part and the ``oil-loving'' tail part of each 
surfactant is assumed to be identical to a water molecule and an oil 
molecule, respectively. A fraction of the nearest-neighbour pairs of 
dissimilar molecules are connected by a rigid bond; each of the dumb 
bell-like structures where a rigid bond connects two dissimilar molecules 
at its two ends is identified as a surfactant. Thus, in this model, the 
molecules of oil and water occupy the lattice sites whereas the surfactants 
are located on the bonds. The Hamiltonian for this model is given by 
\begin{equation}
{\cal H} = (J/2) \sum (1 - S_i S_j) 
\end{equation}
\begin{figure}
\centerline{\psfig{file=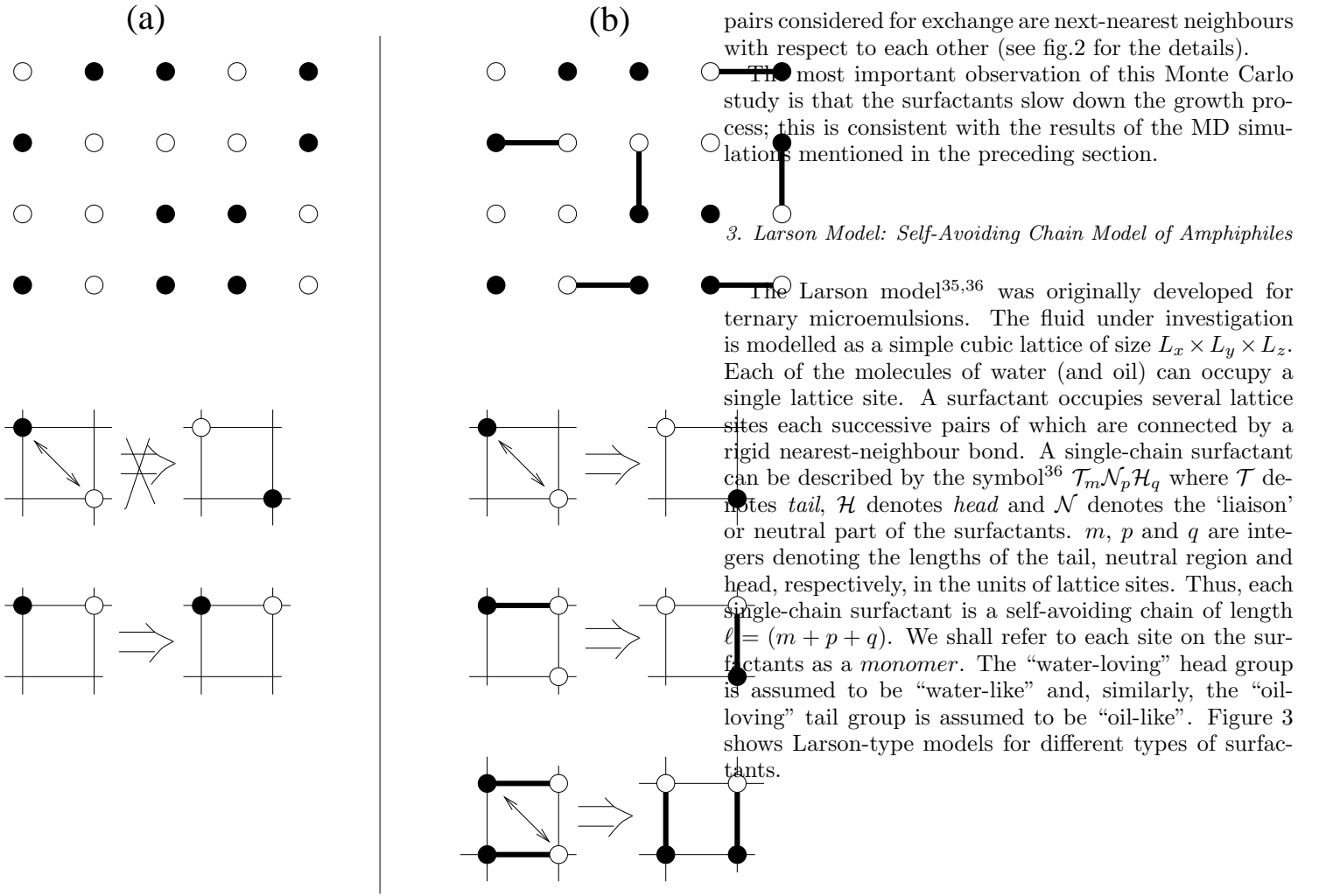,width=12cm}}
\caption{\sf Schematic representations of (a) the lattice model
of a binary alloy and (b) the Kawakatsu-Kawasaki lattice model of 
a ternary system where one of the components is a surfactant. The
elementary `exchange' processes allowed in the dynamics of the
two models are also shown. Exchange of anti-parallel nearest
neighbours only is allowed in the case of the binary alloy whereas
exchange of antiparallel next-nearest neighbours only is allowed
in the case of ternary system}
\label{fig2}
\end{figure} 
Note that the energy of the water-water, oil-oil and water-oil interactions 
are given by $E_{ww} = 0 = E_{oo}$ and $E_{wo} = J$. Since $J > 0$, 
water and oil tend to phase separate in the absence of surfactants. Since 
the two ends of the surfactants are not allowed to split into two 
separate water and oil molecules during the time evolution of the system 
the interaction energy between the water-like end and oil-like end of  
such a surfactant does not affect the dynamics. However, the dynamics 
must distinguish between individual water (and oil) molecules and  the 
water-like (and oil-like) ends of surfactants so as not to split a 
surfactant into a water molecule and a oil molecule. The crucial feature 
of the Kawakatsu-Kawasaki dynamics, which distinguishes it from the 
Kawasaki spin-exchange dynamics for binary alloys, is that the anti-parallel 
spin pairs considered for exchange are next-nearest neighbours with respect 
to each other (see fig.2 for the details). 

The most important observation of this Monte Carlo study is that the 
surfactants slow down the growth process; this is consistent with 
the results of the MD simulations mentioned in the preceding section.  

\subsubsection{Larson Model: Self-Avoiding Chain Model of Amphiphiles} 

The Larson model \cite{lar,livrev}
 was originally developed for ternary microemulsions.  
The fluid under investigation is modelled as a simple cubic lattice of size 
$L_x \times L_y \times L_z$. Each of the molecules of water (and oil) can  
occupy a single lattice site. A surfactant occupies several lattice sites 
each successive pairs of which are connected by a rigid nearest-neighbour bond. 
A single-chain surfactant can be described by the symbol \cite{livrev}
${\cal T}_m{\cal N}_p{\cal H}_q$ where ${\cal T}$ denotes {\it tail}, 
${\cal H}$ denotes {\it head} and ${\cal N}$ denotes the `liaison' or neutral 
part of the surfactants. $m$, $p$ and $q$ are integers denoting the lengths 
of the tail, neutral region and head, respectively, in the units of lattice 
sites. Thus, each single-chain surfactant is a self-avoiding chain of 
length $\ell = (m+p+q)$. We shall refer to each site on the surfactants as a 
$monomer$. The ``water-loving'' head group is assumed to be ``water-like'' 
and, similarly, the ``oil-loving'' tail group is assumed to be ``oil-like''. 
Figure 3 shows Larson-type models for different types of surfactants.  
\begin{figure}
\centerline{\psfig{file=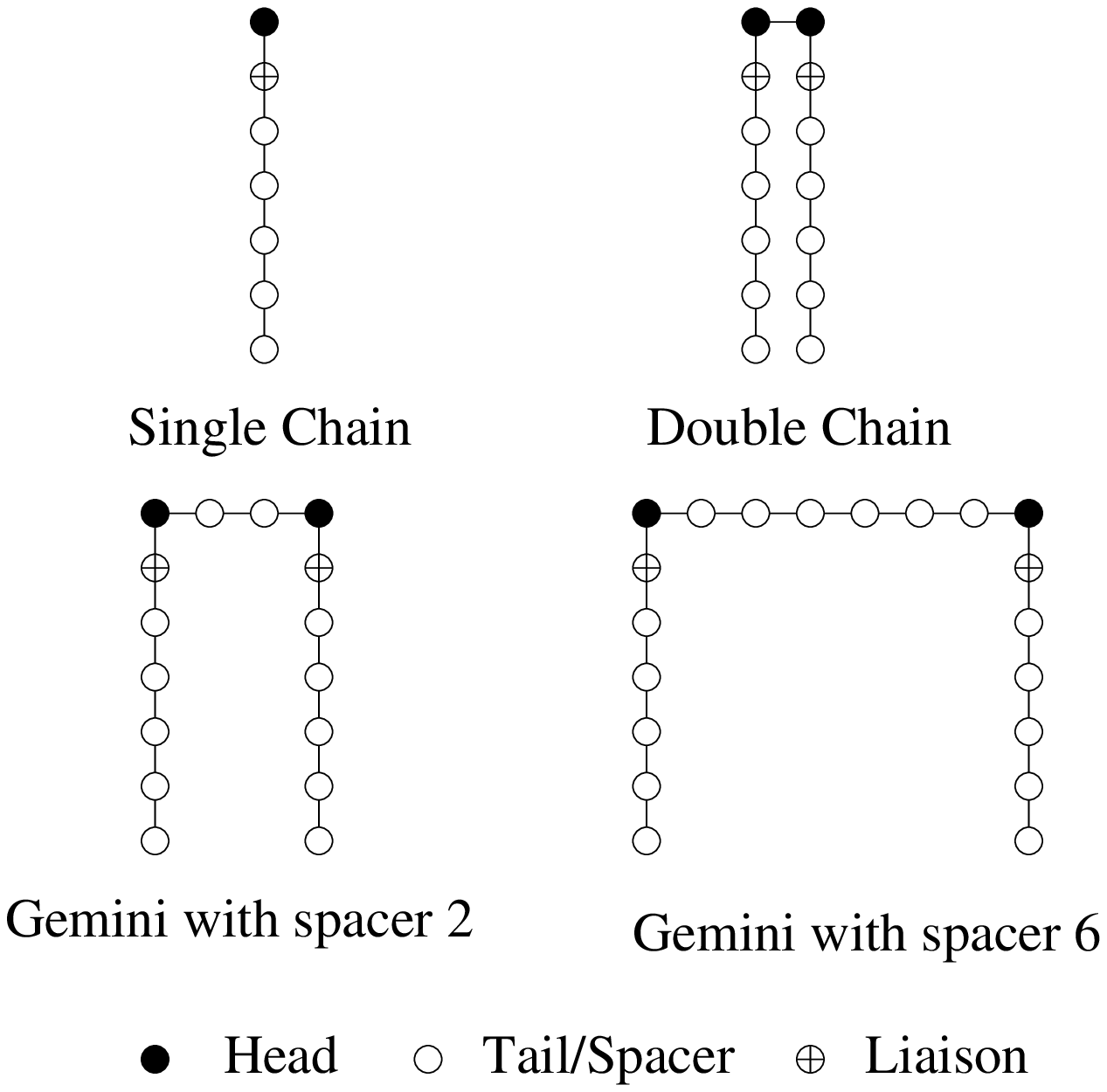,width=12cm}}
\caption{\sf Larson-type models of single-chain, double-chain
and gemini surfactants.}
\label{fig3}
\end{figure}
 
Jan, Stauffer and collaborators \cite{st} simplified the Larson model 
by formulating it in terms of Ising-like variables 
in the same spirit in which a large 
number of simpler lattice models had been formulated earlier \cite{gomp} 
for the convenience of calculations. In this reformulation, a classical Ising 
spin variable $S$ is assigned to each lattice site; $S_i = 1$ ($-1$) if the 
$i$-th lattice site is occupied by a water (oil) molecule. If the $j$-th 
site is occupied by a monomer belonging to a surfactant then $S_j = 1, -1, 0$ 
depending on whether the monomer at the $j$th site belongs to head, tail or  
neutral part. The monomer-monomer interactions are taken into account through 
the interaction between the corresponding pair of Ising spins which is assumed 
to be non-zero provided the spins are located on the nearest-neighbour sites 
on the lattice. Thus, the Hamiltonian for the system is given by the 
standard form 
\begin{equation}
H = - J \sum_{<ij>} S_i S_j. 
\end{equation}
where attractive interaction (analogue of the ferromagnetic interaction in 
Ising magnets) corresponds to $J > 0$ and repulsive interaction (analogue 
of antiferromagnetic interaction) corresponds to $J < 0$ \cite{st}. 
The temperature $T$ of the system is measured in the units of $J/k_B$  
where $k_B$ is the Boltzmann constant. The Kawakatsu-Kawasaki model 
may be regarded as a special case of the Larson model, namely, 
$m = q = 1$, $p = 0$.

Jan, Stauffer and collaborators \cite{st} extended the model further to 
describe single-chain surfactants with {\it ionic} heads. According to their 
formulation, the monomers belonging to the ionic heads have Ising spin 
$+2$ to mimic the presence of electric charge. The repulsive interaction 
between a pair of ionic heads is taken into account through an 
(antiferromagnetic) interaction $J = -1$ between pairs of nearest neighbour 
sites both of which carry spins $+2$; however, the interaction between all 
other pairs of nearest-neighbour spins is assumed to be $J = 1$. By 
restricting the range of the repulsive (antiferromagnetic) interaction 
between the ``charged'' heads to only one lattice spacing one is, effectively, 
assuming very strong screening of the Coulomb repulsion between ionic heads 
by the counterions. 

Starting from an initial state (which depends on the phenomenon under  
investigation), the system is allowed to evolve following the standard 
Metropolis algorithm: each of the attempts to move a surfactant takes 
place certainly if $\Delta E < 0$ and with a probability proportional to 
$exp(-\Delta E/T)$ if $\Delta E \geq 0$, where $\Delta E$ is the change 
in energy that would be caused by the proposed move of the surfactant 
under consideration.

\begin{figure}
\centerline{\psfig{file=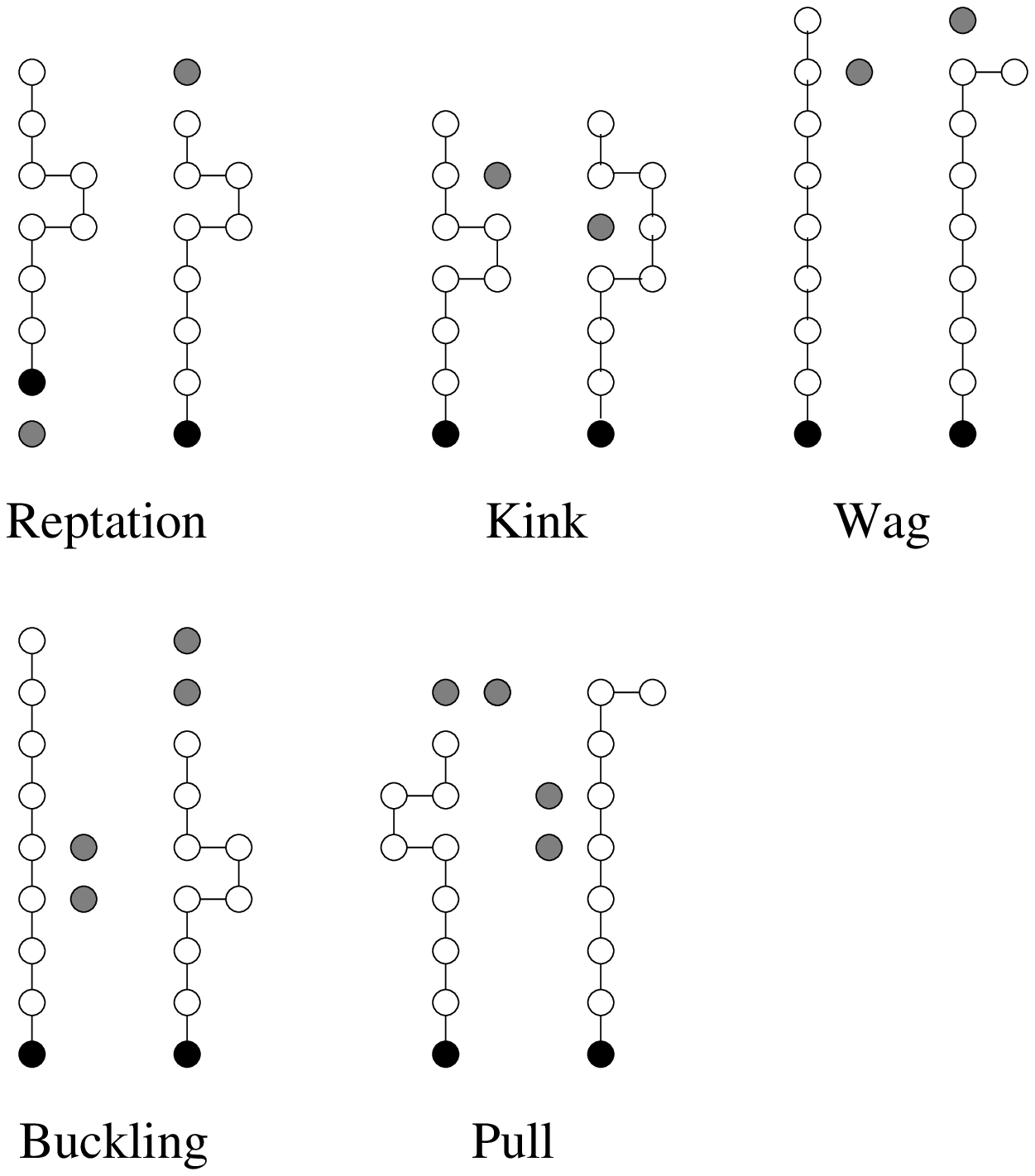,width=12cm,height=12cm}}
\caption{\sf Different moves of the amphiphiles} 
\label{fig4}
\end{figure}

Next, we specify the allowed moves of the  surfactants for the appropriate 
sampling of the states of the system in a MC simulation. So far as the 
single-chain surfactants are concerned, the moves allowed for the 
surfactants are as follows: (see fig. 4)\\
(i) {\it reptation:} one of the two ends of each surfactant is picked up 
randomly, with equal probability, and the surfactant is allowed to move 
forward along its own contour by one lattice spacing with the probability 
mentioned above; this move effectively mimics the reptile-like slithering of 
the surfactants and hence the name;\\ 
(ii) {\it spontaneous chain buckling:} a portion in the middle of the tail 
is randomly picked up and allowed to buckle with the probability mentioned 
above;\\
(iii) {\it kink movement:} a kink formed by the buckling or reptation 
is allowed to move to a new position with the appropriate probability 
calculated according  to the prescription mentioned above;\\
(iv) {pull move:} this is the reverse of spontaneous chain buckling; a 
buckled part of the tail is pulled so as to make it more extended. 
Each of these moves is possible only 
if the new positions of all the monomers are not occupied simultaneously 
by monomers belonging to other surfactants. Each surfactant is allowed 
to try each of the above mentioned moves once during each MC step.   

Note that the monomers of the same surfactant as well as different 
surfactants are not allowed to occupy the same lattice site simultaneously; 
this represents a hard-core intra-chain as well as inter-chain repulsion 
for monomer separations smaller than one lattice spacing. Moreover, at 
any non-vanishing temperature, during the out-of-line thermal fluctuations 
of the chains, the hard-core repulsion leads to steric repulsion 
between the chains. To our knowledge, no potential energies associated 
with the torsion of the surfactant chains have been incorporated so far in 
any work on Larson-type models. 

A microscopic lattice model of double-chain surfactants (with a single 
head) in aqueous solution was developed by Bernardes \cite{bern2} by 
modifying the Larson model of single-chain surfactants \cite{lar,st,livrev}. 
In terms of the symbols used above to denote the primary ``structure'' of 
the microscopic lattice model of single-chain surfactants, Bernardes' 
lattice model of double-chain surfactants, with a single hydrophilic head, 
can be described by the symbol 
${\cal T}_m{\cal N}_p{\cal H}_q{\cal N}_p{\cal T}_m$.   
Very recently, Maiti and Chowdhury \cite{maiti}
 have proposed a microscopic lattice 
model of gemini surfactants by extending Bernardes' model so as to 
incorporate two hydrophilic heads connected by a {\it spacer} . 
This model of a gemini surfactant can be represented by the symbol 
${\cal T}_m{\cal N}_p{\cal H}_q{\cal S}_n{\cal H}_q{\cal N}_p{\cal T}_m$ 
where $n$ is the number of lattice sites constituting the spacer represented 
by the symbol ${\cal S}$ and the other symbols have the same meaning 
as in the case of Bernardes' model of double-chain surfactants (see fig.3). 
For the convenience of computation, the Bernardes' model of double-chain 
surfactants as well as Maiti and Chowdhury's model of gemini surfactants 
have also been formulated in terms of classical Ising spin variables, 
generalizing the corresponding formulation for the single-chain surfactants 
reported in  ref \cite{st}. 

To our knowledge, the work of Bernardes et {\it al.}\cite{bern96}
 is the only 
published report of the investigation on the  
kinetics of phase ordering in the Larson model of microemulsions.
Oil, water and surfactant molecules were distributed randomly in 
the initial state of the system. The system was allowed to evolve 
by implementing the moves of the surfactants according to the 
algorithm mentioned earlier while the molecules of oil and water 
were allowed to exchange their positions according to the Kawasaki-spin 
exchange dynamics mentioned earlier in the context of binary mixtures. 
The time evolution of the mean sizes of the clusters of oil, water 
and amphiphiles in the system were observed up to  200,000 MC steps. 

When the amphiphile concentration $\phi_a = 0.1$ 
and the concentrations of oil and water $\phi_o = \phi_w = 0.45$, 
respectively, the average length scale 
$<R>_{ow}=[(1/2)(\chi_o + \chi_w)]^{1/3}$ was found to obey 
the power law $<R>_{ow} \sim t^n$ with $n \simeq 1/3$, i.e., the coarsening 
was found to 
be governed by the same Lifshitz-Slyozov law which governs the coarsening 
in binary mixtures in the absence of amphiphiles. However, comparing 
these results with the corresponding results obtained by other groups 
using various different techniques, we believe that the truly 
asymptotic logarithmically slow growth regime lies beyond the longest 
time scales of observation in the computer experiments of Bernardes et 
al.

The kinetics of ordering in microemulsions and micellar 
solutions containing gemini surfactants have not been reported so far. 
A microscopic model for single-chain surfactants at the air-water 
interface has been developed earlier by one of us \cite{Chow3,Chow4} by 
appropriately modifying the Larson model \cite{lar,st,livrev} of 
ternary microemulsions \cite{chen,Gelbert}. Later Maiti and Chowdhury 
replaced the single-chain surfactants in the model introduced in ref.
\cite{Chow3} by the model gemini surfactants , thereby 
getting the desired microscopic model of gemini surfactants at the 
air-water interface. A novel entropy-driven phase segregation has 
been observed in computer experiments on a binary mixture of chemically 
identical single-chain surfactants of two different lengths at the 
air-water interface. However, its dynamical aspects have not been 
investigated so far.

\subsection{Hybrid Models} 

In this model the binary mixture of $A$ (say, oil) and $B$
(say, water) is represented by a continuum field $\psi(\vec r)$
in the same fashion as outlined in section 2 in the context 
of the continuum description of binary alloys. However,
unlike the continuum model discussed in the preceding
section, the amphiphiles are treated at the molecular level
by describing the dynamics of the positions of the center
of gravity and the orientations of the molecules. That
is why this model is called ``hybrid''.

In this model the surfactant molecules are modelled as ``dumbbells''
of length $l$ which have two interactions centers at the two
ends, one of which is $A$-philic and the other is $B$-philic.
Kawakatsu and Kawasaki \cite{kawasaki} assumed that the ``$A$-philic
and $B$-philic interactions centers of the surfactant have
the same chemical species as $A$ and $B$ components of the
binary mixture, respectively''. Suppose, the position of the
center of gravity of the $i$-th surfactant molecule is denoted
by $\vec r_i$ and the unit vector from the $B$-philic interaction
center to the $A$-philic interaction center of the same molecule
by the symbol $\hat s_i$. For simplicity, also assume that
$V_{AA}(\vec r) = V_{BB}(\vec r) = \phi(\vec r)$ and 
$V_{AB}(\vec r) = \chi(\vec r)$. Then, the free energy functional
is given by 
\begin{equation}
F = F_{\psi \psi} + F_{\psi s} + F_{ss}
\end{equation}
where
\begin{equation}
F_{\psi \psi} =\int d^dr [(1/2) C_{\psi}\{\nabla\psi(\vec r)\}^2
+(-r_o)\psi^2(\vec r) + u\psi^4(\vec r)]
\end{equation}
\begin{equation}
F_{\psi s} =\mu_s N_s + (ql/2)\sum \int d^d r V_{-}(\vec r - 
\vec r_i)\hat s_i {\bf .} \nabla\psi(\vec r)
\end{equation}
$$F_{ss}  =  q^2\sum[2V_+(r_{ij}) + 
(l^2/4)(s_{ij}^-s_{ij}^-)\nabla\nabla\phi(r_{ij})$$
\begin{equation}
  +  (l^2/4)(s_{ij}^+s_{ij}^+)\nabla\nabla\chi(r_{ij})]
\end{equation}
with
\begin{equation}
V_{\pm}(r) = \phi(r) \pm \chi(r)
\end{equation}
\begin{equation}
s_{ij}^{\pm} = \hat s_i \pm \hat s_j
\end{equation}
\begin{equation}
r_{ij} = |\vec r_i - \vec r_j|
\end{equation}
\begin{equation}
\mu_s = q\rho\int d^dr V_+(\vec r)
\end{equation}
where
\begin{equation}
\rho = \rho_A(\vec r) + \rho_B(\vec r)
\end{equation}
is assumed to be a constant.
The equations of motion are given by 
\begin{equation}
\partial\psi(\vec r,t)/\partial t = L_{\psi} \nabla^2[\delta F/\delta\psi]
\end{equation}
\begin{equation}
d\vec r_i(t)/dt = -L_{\rho}(\partial F/\partial\vec r_i)
\end{equation}
\begin{equation}
d\hat s_i(t)/dt = - L_s[(\partial F/\partial\hat s_i) -
\{(\partial F/\partial \hat s_i){\bf .}  \hat s_i \} \hat s_i]
\end{equation}
where $L$'s are phenomenological kinetic coefficients. The 
second term on the right hand side of the Eq. (29) arises 
from the constraint $|\hat s_i| = 1$. The Eqs. (27)-(29)
were solved numerically assuming the forms 
$\phi(r) = -\exp(-r) $ and $ \chi(r) = \alpha \exp (-r)$.
Equation (27) was solved by the cell dynamic method
whereas  Eqs. (28) and (29) were solved by the molecular
dynamic method (see the original paper for the numerical
values of the various parameters). Two kinds of initial
conditions were used: (a) in case of equal volume fractions
of the $A$ and $B$ components a random bicontinuous structure
formed and coarsened with time (b) in the case where the
volume fraction of the $B$ component were three times that 
of the $A$ component a dispersion of droplets was formed 
which coarsened with time. 

In both the situations (a) and (b) mentioned above, Kawakatsu 
and Kawasaki \cite{kawasaki3} found a crossover in the temporal 
evolution of $R(t)$ from the power law 
$R(t) \sim t^{1/3}$ to a slower 
growth. Such a crossover was found to occur 
when the oil-water 
interface gets saturated by the surfactants.  

Kawakatsu et {\it al.} \cite{kawakatsu95}
have also studied phase separation in 
immiscible binary mixtures in the presence of surfactant molecules 
of asymmetric shape using the hybrid model. They observed a spontaneous 
morphological change from a bicontinuous structure to a micellar 
structure during the phase separation process which gives rise to 
the formation of a amphiphilic monolayer at the oil-water interface. 
This is consistent with the known fact that the asymmetry of molecular 
shape gives rise to spontaneous curvature of amphiphilic monolayers 
\cite{helfrich,lipo,peliti,hel1,nelson,safran}.

\subsection{Purely Phenomenological Models}

This model is an extension of the continuum model of binary
alloy phase ordering. Naturally, this is an appropriate extension  
of Eqs. (2)-(4) above. Suppose $\rho(\vec r)$ is the
density of the surfactant molecules at the location $\vec r$.
Laradji et {\it al}. \cite{larad2,larad3} postulated that the 
effective free energy functional for the ternary system under
consideration is given by
$$F[\psi(r),\rho(r)] = \int d^dr[c(\nabla \psi)^2 -r_o\psi^2 + u\psi^4$$
\begin{equation}
\quad \quad \quad \quad \quad \quad \quad \quad  + g\rho^2\psi^2 + a\rho^2 -\mu\rho - (\Delta\mu)\psi + F_s
\end{equation}

Here $\rho(r)$ is the density of the surfactant molecules at the location
$r$. $c, r_o, u, g, a$ and $ \mu$ are phenomenological coefficients.
$\mu$ is the chemical potential of the amphiphilic molecules.
$\Delta\mu$ is the difference of the chemical potentials of water and
oil. Note that $g$ is the strength of the coupling between
the two fields $\rho$ and $\psi$ and Eq.(30) is an extension
of equation (2).
The equations of motion in this model are given by
\begin{equation}
\partial\psi/\partial t = \nabla^2(\delta F/\delta \psi)
+\eta_{\psi}(\vec r,t)
\end{equation}

\begin{equation}
\partial\rho/\partial t = \nabla^2(\delta F/\delta\rho)
+ \eta_{\rho}(\vec r,t)
\end{equation} 
where 
\begin{equation}
<\eta(\vec r,t)\eta(\vec r^{\prime},t^{\prime}) =
2k_BT\nabla^2\delta(\vec r -\vec r^{\prime})\delta(t-t^{\prime})
\end{equation}
The above equations (31)-(33) are the generalizations of the
Eqs. (3) and (4). In the terminology of the Hohenberg-Halperin
classification scheme \cite{hohen} this model corresponds to the
model D. The surfactant property of the amphiphiles is taken 
into account through
\begin{equation}
F_s = s\int d^dr \rho(\nabla\psi)^2
\end{equation}
 
where $s$ is a phenomenological constant.
Laradji et {\it al}. \cite{larad} solved the Eqs. (30)-(33)
numerically (see the original paper for the numerical 
values of the parameters) starting from a random
initial condition, i.e., to each grid point $\psi(\vec r)$
and $\rho(\vec r)$ were assigned small random values around
their initial values at $t=0$. The second moment of circularly
averaged structure factor, $R^{-1}(t)$, was monitored as a 
function of time $t$. The main results of their
investigations are as follows:
(i) The location of the peak of the structure factor moves
initially to small $k$ as time passes, thereby indicating
coarsening. However, the coarsening seems to come to a halt
at very late stages because the peak position was observed 
to become static at a fixed $k ~=~k_e \ne 0$. Moreover, the
larger is the concentration of the surfactants, the smaller
is the final size of the oil-rich (or water-rich) domains.
(ii) $R(t) \propto (log t)^y$, although estimation of $y$
was not carried out by these authors. The slow growth
observed in this study is consistent with the corresponding
results of the microscopic models in the preceding section.

Recently, Ahluwalia and Puri \cite{alu} have made an attempt to study the 
kinetics of phase ordering in ternary microemulsions using a CDS approach. 
The results are consistent with those obtained by the other methods 
discussed so far.
\section{Summary and Conclusion}

For the description of those phenomena which are strongly dependent on 
steric interactions, the Larson model is believed to be more realistic 
than the  lattice model of Kawakatsu and Kawasaki. However, if these two 
models belong to the same dynamic universality class, then the long-time 
regime, where Kawakatsu and Kawasaki have observed non-algebraic slow growth,
must be lying beyond the longest runs made by Bernardes et al.\cite{bern96}. 

It is worth mentioning here that the hybrid model takes into
account the surfactant property of amphiphilic molecules in a much
more realistic manner than that in the purely phenomenological models.
Nevertheless, the hybrid  model suffers from the shortcomings that the 
excluded volume of the surfactants is not taken into account because 
of the assumption that $\rho_A(\vec r) + \rho_B(\vec r) = constant$. 

The amphiphilic molecules in ternary microemulsions and micellar solutions 
may be regarded as impurities added to an immiscible binary mixture of oil 
and water. But, these are not "quenched" impurities and, therefore, the 
process of phase ordering is not driven by thermally activated motion of 
the oil-water interface. 
Almost all the numerical works published so far demonstrate that 
during the late stages of the coarsening process the oil-water 
interface is saturated by the surfactants leading to a reduction of 
the oil-water interfacial tension which gives rise to the crossover 
from the Lifshitz-Slyozov-like power-law growth to a slower growth 
of the ordered regions. This is consistent with the results obtained 
by incorporating the effects of the surfactants in the Lifshitz-Slyozov 
theory for the kinetics of phase ordering in immiscible binary 
mixtures \cite{yao}.

Experimental study of the kinetics of ordering in ternary microemulsions 
and micellar solutions is difficult because this phenomenon is too fast to 
observe over time scales which would be sufficient to extract the growth 
laws. $A-B$ diblock copolymer in a $A/B$ binary homopolymer blend behaves 
effectively like a surfactant because the $A$ ($B$) subchain of the copolymer 
dissolves preferentially in the $A$-rich ($B$-rich) phase ~\cite{bates}.
Since the phase separation process in polymer systems is much 
slower because of entanglements of the polymer chains, the kinetics of  
ordering in a binary polymer blend containing an amphiphilic block 
copolymer can be utilized for experimental investigations. 
The results of such investigations can be compared with the theories 
of the kinetics of phase ordering in ternary microemulsions and 
micellar solutions. 
 
\section*{Acknowledgements} 
One of us (DC) thanks A.T. Bernardes, T.B. Liverpool, 
and D. Stauffer for many illuminating discussions on amphiphilic systems and 
the Alexander von Humboldt Foundation for supporting the work on 
amphiphilic systems at IITK through a research equipment grant.  
DC was introduced to the analytical techniques of studying kinetics of 
phase ordering by Jim Gunton twelve years ago; we dedicate this review to 
Jim on his 60th Birthday.

\end{document}